\documentstyle[12pt]{article}

\newcommand{\be}{\begin{equation}}
\newcommand{\ee}{\end{equation}}

\newcommand{\bfR}{{\sf R\hspace*{-0.9ex}\rule{0.15ex}%
{1.5ex}\hspace*{0.9ex}}}
\newcommand{\N}{{\sf N\hspace*{-1.0ex}\rule{0.15ex}%
{1.3ex}\hspace*{1.0ex}}}

\date{\ }
\begin{document}

\title{Skyrme model on $S_3$ and Harmonic maps}

\author{{\large Y. Brihaye}$^{\diamond}$
and {\large C. Gabriel
 \thanks{$^{\dagger}$ Aspirant F.N.R.S}
}$^{\dagger}$
\\
$^{\diamond}${\small Dep. of Mathematical Physics, University of Mons-Hainaut,
Mons, Belgium}\\
$^{\dagger}${\small Dep. of Mechanics and Gravitation, University
of Mons-Hainaut, Mons, Belgium}
}

\date{}

\maketitle
\begin{abstract}
A non-linear sigma model mimicking the Skyrme model on 
$S_3$ is proposed and a family of classical solutions to the
equations  are constructed  numerically. 
The solutions terminate into  catastrophy-like spikes
 at critical values of the Skyrme coupling 
constant and, when this constant is zero, they coincide with the 
series of Harmonic maps on $S_3$ constructed 
some years ago by P. Bizon.
\end{abstract}
\medskip
\medskip

\section{Introduction}
A few years ago P. Bizon \cite{bi} constructed a remarkable sequence of
harmonic maps from $S_3$ into $S_3$. Only a few examples are known of non-linear
equations which admit such a sequence of regular solutions. Some of them occur in the framework of gauged non-linear sigma models.
The prototype of non-linear sigma model
in theoretical physics in the Skyrme model \cite{sk}.
It is based on a mapping of the physical space $\bfR^3$ into
the manifold of the Lie group SU(2), wich is homeomorphic to $S_3$.
In order to guarantee the existence of topological soliton the
physical space is compactified. The original Skyrme model possess
an SU(2)$\otimes$SU(2) global symmetry which can partly be gauged.
Several examples of sequence of regular-finite energy classical
solutions appear in different gauged vesions of the Skyrme model
\cite{boku,brku}. 

For all known examples, an interesting phenomenon,
related to catastrophe theory, occurs to the solutions when the minimal (i.e.
quadratic in the derivatives) sigma model is supplemented by a quartic
term in the derivatives  \cite{eks,brku,bks}, a so called Skyrme term.  
The higher derivative interaction term
is proportional to a free constant, say $\epsilon$,
which constitutes a coupling constant of the model.
To be more concrete let us denote 
(in the models \cite{eks,brku,bks})
by $S_n(\vec x,\epsilon=0), n \in \N$ the
sequence of solutions available in absence of the Skyrme term
and $E_n(0)$ the corresponding classical energy. 
Solving the classical equations for $\epsilon>0$, it appears that
a branch of solutions $S_n(\vec x,\epsilon)$ exists for $0\leq
\epsilon\leq \epsilon_{cr}(n)$ and that no solution is available for
$\epsilon>\epsilon_{cr}(n)$. However, there exists a second branch of 
solution, say
$\tilde S_n(\epsilon)$, for $\epsilon \in ]0,\epsilon_{cr}(n)]$.
The numerical results indictate that the following properties 
hold for $n \in \N$
\be
\tilde E_n(\epsilon) < E_n(\epsilon)
\ee
\be
\lim_{\epsilon\rightarrow \epsilon_{cr}(n)}\tilde E_n(\epsilon)=E_n(\epsilon)
\ee
\be
\lim_{\epsilon\rightarrow 0} \tilde E_n(\epsilon) = E_{n-1}(\epsilon)
\ee
In fact the statement (2) is even stronger because we have in fact 
\be
\lim_{\epsilon\rightarrow \epsilon_{cr}(n)} S_n(\vec x, \epsilon) = 
\lim_{\epsilon\rightarrow \epsilon_{cr}(n)} \tilde S_n(\vec x, \epsilon) =
S_n(\vec x, \epsilon_{cr}(n))
\ee
in a uniform sense for $\vec x \in \bfR^3$. 
The counterpart of Eq.(4) related to the limit (3)  does not hold
because the solutions $S_n(\vec x,\epsilon)$ and
$S_{n-1}(\vec x,\epsilon)$ do not in general obey the same boundary conditions.

The plot of the energies $E(n,\epsilon)$ and $\tilde E(n,\epsilon)$ as a
function of $\epsilon$ shows that the two curves terminate at $\epsilon
= \epsilon_{cr}(n)$ into a cusp, characteristic of catastrophe theory.

\par 
This phenomenon is absent in the normal (i.e. ungauged) Skyrme model but
it occurs in several gauged versions of this model, namely in the 
``hidden gauge'' Skyrme model \cite{jap}.
In passing let us point out that the trivial topological sector
sector of the ``hidden gauge'' Skyrme model is equivalent to the
(bosonic part of the) electroweak
standard model considered in the limit of an infinitely heavy Higgs particule
(i.e. for $M_H \sim \lambda \rightarrow \infty$) \cite{eks,brku}.

\par Owing the above properties associated with sequences of classical
solutions available in the gauge-Skyrme model, it is natural to study
how the harmonic maps of Ref \cite{bi} respond to the addition of
a Skyrme interaction term.
This is the aim of this paper.

\section{The model}
The functional energy studied in \cite{bi} reads
\be
\label{Ephi}
E_0 (f) = \int_{S_3} h_{AB}(f){\partial f^A\over{\partial x^i}} 
{\partial f^B\over{\partial x^j}}  g^{ij}dV
\ee
where $f$ represents a mapping from $S_3$ (with local coordinates noted
$x^i$) into $S_3$ (with local coordinates note $f^A$). We use the same
notations as in \cite{bi}, namely $x^i = (\psi,\theta,\phi)$ are the standard
coordinates on $S_3$ with the metric
\be
\label{metric}
ds^2 = d\psi^2 + \sin^2\psi(d\theta^2 + \sin^2 \theta d\phi^2)
\ee
and  similarly $f^A = ( \Psi, \Theta, \Phi)$ on the $S_3$ target space. 

The functional $E_0(f)$ is  a non-linear sigma model; the
classical solutions, i.e. the extremum of the functional,
are called harmonic
maps in mathematics. In this paper, we study the extremum of the
extended energy functional
\be
E(f) = E_0(f) + E_{Sk}(f)
\ee
with the Skyrme interaction term
\be
\label{Esk}
E_{sk}(f) = \int_{S_3} h_{AB}(f) h_{CD}(f){\partial f^A\over{\partial x^i}}
{\partial f^B\over{\partial x^j}}{\partial f^C\over{\partial x^k}}
{\partial f^D\over{\partial x^l}}(\mu_1g^{ij}g^{kl}+ \mu_2g^{il}g^{jk})dV
\ee
where $\mu_1,\mu_2$ are constants. In order to keep the model (\ref{Esk})
as close as possible to the standard Skyrme model defined on $\bfR\ ^3$
we choose $\mu_1 = -\mu_2 \equiv \kappa^2$.
\par Finally, we also assume that the function $f^A(x^i)$ is equivariant
i.e. invariant under the SO(3) rotations corresponding to the angles 
$\theta$ and $\phi$. This leads us to the ansatz
\be
\Psi(\psi) = f(\psi) \quad , \quad \Theta = \theta \quad , \quad  \Phi = \phi
\ee
in terms of which the functional $E(f)$ reads
\be
\label{ESS}
E(f) = 4\pi \int^{\pi}_0 \lbrace \sin^2\psi 
({df \over d \psi})^2 + 2 {\sin^2 f \over \sin^2\psi} 
+ \kappa^2({\sin^2f \over \sin^2\psi}
 ({df \over d \psi})^2 + {\sin^4f\over{2\sin^4\psi}})\rbrace d\psi
\ee
The classical equations are obtained by varying $E(f)$ with respect to $f$.
We are interested by solutions such that the energy density
(i.e. the term in bracket in the integral above) is regular.
This imposes in particular $f(0) = f(\pi) = 0$.

In the purpose of comparison with the normal Skyrme model on $\bfR^3$
we just write the expression of the classical energy in this model:
\be
\label{Eskyrme}
E(f) = 4\pi \int^{\infty}_0  r^2 \lbrace 
({df \over dr})^2\  + 2 {\sin^2 f \over r^2}
+ \kappa^2({\sin^2f \over r^2} ({df \over dr})^2 
+ {\sin^4f\over{2 r^4}})\rbrace dr
\ee

\section{The numerical solutions}
It is convenient to label the extrema of the functional
(\ref{ESS}) by $f_n(\psi,\kappa^2)$, where the
integer $n$  counts the number of times the solution
crosses the line $f={\pi\over 2}$.
It is also useful to notice the discrete 
invariances of the equation~:
\be
\label{sym}
     f(\psi) \rightarrow f(\pi - \psi) \quad , \quad
     f(\psi) \rightarrow \pi - f(\pi - \psi) \ \ \ .
\ee
They transform any  solution into another
one with the same energy.

In \cite{bi} a sequence of extrema 
$f_n(\psi, 0)$ (also called critical maps)  to the functional (\ref{Ephi})
 was constructed.
It is such that $f_0(\psi,0) = 0$ , $f_1(\psi,0) = \psi$. 
To our knowledge,
the solutions corresponding to $n>1$ cannot be expressed
in terms of known functions.
The profiles of the functions $f_2(\psi,0)$ and $f_3(\psi,0)$ 
are represented on Fig.1 and Fig.2 respectively.
For the first few values of $n$, the energies 
are given \cite{bi} in the  table below~:

\vspace{0.5cm}
\begin{center}
\begin{tabular}{|l|l|}
\hline
n &$E(n)/6\pi^2$\\
\hline
0 &0\\
1&1\\
2&1.2319\\
3&1.3036\\
4 &1.3240\\
\hline
\end{tabular}
\end{center}
\vspace{0.5cm}

We now discuss the evolution of these solutions for $\kappa^2 \neq 0$.
Obviously the vacuum solution $f_0=0$ 
(or, by (\ref{sym}), $f_0=\pi$)  exits
for all values of $\kappa^2$ and its energy is zero. 
The solution corresponding to $n=1$ reads
\be
\label{psi1}
f_1(\psi,\kappa^2)=\psi \qquad {\rm or}  \qquad \ f_1(\psi,\kappa^2)=\pi-\psi
\ee
and, like the vacuum,
also exists irrespectively of the constant $\kappa^2$. It has an energy
\be
E_1(\kappa^2) = 6\pi^2(1+{\kappa^2\over 2})
\ee

Solving numerically the equations for finite values of $\kappa^2$ 
we find that the solution $f_n(\psi,0)$ gets continuously deformed 
and generates a branch of solutions $f_n(\psi,\kappa)$,
at least for small values of $\kappa$.
For  $n=2$ the solution exists on the interval
\be
\label{int2}
0\leq \kappa^2 \leq \kappa^2_{cr}(2)\quad ;\quad \kappa^2_{cr}(2)\simeq
0.261
\ee
\be
E_2(\kappa^2_{cr}(2))/6\pi^2 \simeq 1.670
\ee
The profile of $f_2(\psi,\kappa_{cr}^2(2))$ is also shown on Fig.1
and the energy of the solution is represented (in function of $\kappa$)
by the upper part of the $n=2$ curve in Fig.3.
We find no solution  for $\kappa^2 > \kappa_{cr}^2$ but 
a second branch of solutions, say $\tilde f_2(\psi, \kappa^2)$,
exists on the interval  (\ref{int2}). 
For fixed $\kappa$,
the energy of the solution $\tilde f_2$ 
is lower than the energy of $f_2$,
i.e. (using an obvious notation)
\be
     \tilde E_2 (\kappa) \leq E_2 (\kappa)
\ee
as  represented by the lower and upper branches of 
the $n=2$ curve in Fig.3.  At $\kappa^2  = \kappa^2_{cr}(2)$
the  solutions  $f_2$ and $\tilde f_2$ coincide and the
energy plot terminates in a cusp shape, 
suggesting some catastrophic behaviour.

We would like to stress that the evolution of the solution 
profile from the upper branch to the lowest one is completely
smooth, the profiles of the solution 
$\tilde f_2(\psi,0.2), \tilde f_2(\psi, 0.002)$ 
are displayed of Fig.1.
The numerical analysis  strongly suggests
(as illustrated by Fig.1) that
\be
\lim_{\kappa^2\rightarrow 0} \tilde f_2(\psi, \kappa^2) = \pi \quad 
{\rm for} \quad  0<\psi<\pi
\ee
This statement is also supported by the evaluation of the energy 
\be
\lim_{\kappa^2\rightarrow 0} \tilde E_2(\kappa^2)=E_0(0)=0. 
\ee
as indicated  on Fig.3 

\par 
Similarly, the branch developping from the $n=3$ solution exists on the interval
\be
0\leq \kappa^2 \leq \kappa^2_{cr}(3)\quad ; \quad 
\kappa^2_{cr}(3) \simeq 0.017
\ee
and
\be
E_3( \kappa^2_{cr}(3))/6\pi^2 \simeq 1.395
\ee
Again the second branch of solutions $\tilde f_3(\psi,\kappa^2)$ occurs and a few
of these solutions are superposed on Fig. 2 
(resp. $f_3(\psi,0),
f_3(\psi, \kappa^2_{cr}(3))$, $\tilde f_3(\psi,0.015)$,
$ \tilde f_3(\psi, 0.0005))$. The numerical results suggest that
\be
\lim_{\kappa^2\rightarrow 0} \tilde f_3(\psi,\kappa^2) = \pi - \psi \qquad 
{\rm for}
\qquad 0 < \psi<\pi
\ee
to be compared with (\ref{psi1}).
We also find $\lim_{\kappa^2\rightarrow 0}E_3(\kappa^2) =  E_1(0) =6\pi^2$.

\par 
In view of these results,
 we conjecture that two branches of solutions say, $f_n(\kappa^2)$
and $\tilde f_n(\kappa^2)$, exist for $0<\kappa^2<\kappa^2_{cr}(n)$ and that
the result suggested by our numerical analysis holds for $n>3$, that is to say
\be
\lim_{\kappa\rightarrow 0} \tilde f_n(\kappa^2) = f_{n-2}(0)\quad 
{\rm for} \quad
0<\psi<\pi
\ee
\be
\lim_{\kappa\rightarrow 0} \tilde E_n(\kappa^2) = E_{n-2}(0)
\ee

\section{Stability}
The stability of the solutions constructed in the previous section can be
studied by perturbating the classical solution, say $\bar f(\psi)$. In this
paper, we limit ourselve  to fluctuations depending only on $\psi$, i.e.
\be
\label{flu}
f(\psi) = \bar f(\psi) + \eta(\psi)
\ee
Expanding the functional (10) in powers of $\eta$ leads to a quadratic
form which can be diagonalized. The relevant eigenvalue equation reads
\begin{eqnarray}
\label{mode}
\biggl[ 
&-& (1+\kappa^2
   {\sin^2\bar f\over{\sin^2\psi}}) {d^2\over{d\psi^2}} \nonumber \\
&-& (2 \cot \psi - \kappa^2 \bar f'{\sin 2\bar f\over{\sin^2\psi}})
{d\over {d\psi}} \nonumber\\
&+& \Bigl(  2{\cos 2\bar f\over{\sin^2\psi}} + \kappa^2(\bar f')^2
{\cos 2\bar f\over{\sin^2\psi}} \nonumber \\
&+&{\kappa^2\over 2} {1\over{\sin^4\psi}}
(\sin^2 2\bar f + 2\sin^2\bar f\cos 2\bar f) \Bigr)
\biggr] \eta = \omega^2\eta
\end{eqnarray}
and determines the normal modes $\eta(\psi),\omega^2$ about the
classical solution $\bar f(\psi)$.
\par For the vacuum, $\bar f(\psi)=0$, this equation becomes independent
of $\kappa^2$ and can be solved explitely; the normal modes read 
\be
\label{geg0}
\eta_k(x) = \sin \psi \  C^2_k(\cos \psi) \ \ , \ \  \omega^2_k = 3+k (k+4),
\ee
with $k=0,1,2,\dots$ and
where $C^2_k(t)$ denote the Gegenbauer polynomials
\be
C^2_0=1 \ \ , \ \ C^2_1(t) =t,
\ee
\be
(n+2) C^2_{n+2}(t) = 2(n+3)t C^2_{n+1}(t) - (n+4) C^2_n(t)
\ee

The lowest eigenvalues associated with the solutions $\bar f(\psi) = f_n(\psi,0)$
are given  in \cite{bi}. Apart from the case $n=1$ whose eigenvector have
the form (\ref{geg0}) with eigenvalues
\be
\label{geg1}
\omega^2_k = -1+k(k+4)
\ee
the spectrum has to be computed numerically. In particular the solution
$f_n(\psi,0)$ possesses $n$ independent modes of instability of the form
(\ref{flu}).

We studied Eq. (\ref{mode}) numerically for $\kappa^2>0$ and for
the two modes of lowest eigenvalues associated with the solutions corresponding
to  $n=1$ and $n=2$. Our results are reported on Fig. 4. The
solid lines $a,b$ (resp. $c,d$) represent the modes (\ref{geg0}) , (resp. (\ref{geg1}))  for $k=0,1$; we
see that the negative mode about $f_1(\psi,\kappa^2)$ becomes positive for
$\kappa>0.8$
(this is not sufficient, however, to claim that $f_1(\psi_1\kappa^2)$ is 
a local minimum of (7), general fluctuations have to be considered as well).
We also notice that the solution $\tilde f_2(\psi, \kappa^2)$
has no directions of instability in the channel (\ref{flu}).

\par The two lowest modes relative to the solution $f_2(\psi, \kappa^2)$ (resp.
$\tilde f_2(\psi,\kappa^2)$) are represented by the dashed (resp. dotted)
lines on Fig. 4. The numerical results strongly support the fact that the
lowest modes about the solution $f_2(\psi, \kappa^2)$ and the solutions
$\tilde f_2(\psi, \kappa^2)$ vanish in the limit
$\kappa^2\longrightarrow \kappa^2_{cr}(2)$
This confirm on an explicit example a general expectation (based on
catastrophe theory; see. e.g. \cite{eks}) that the solution on the higher
branch (i.e. $\tilde f_2$) has one more directions of instability than the
solution on the lower branch (i.e. $f_2$).
\section{Conclusions}
The non-linear sigma model based on the mapping of $S^3$ into $S^3$ admit a
rich set of classical solutions \cite{bi}. By supplementing it by a
Skyrme-like term, we have shown that the solutions can be interpolated by
the extrema of the family of models indexed by the coupling constant $\kappa^2$.
\par The energy of the classical solution then leads to a serrated pattern
summarized by Fig. 3. Apart from the two lowest branches, which respectively
play the role of the vacuum ($E=0$) and of the Skyrmion solution of standard
Skyrme model, the other solutions cease to exist at critical values of the
parameter $\kappa^2$, where two branches of solutions terminate into a catastrophic
cusp.
\par The analysis of the normal modes about the different solutions indicates
that their stability strongly depends on the Skyrme term and suggests that
the spectrum of the quadratic fluctuation operator varies continuously with
the parameter  $\kappa^2$.

{\bf Acknowledgements} 
We gratefully acknowledge 
discussions with P. Bizon.

\centerline{Figure Captions}
\begin{itemize}
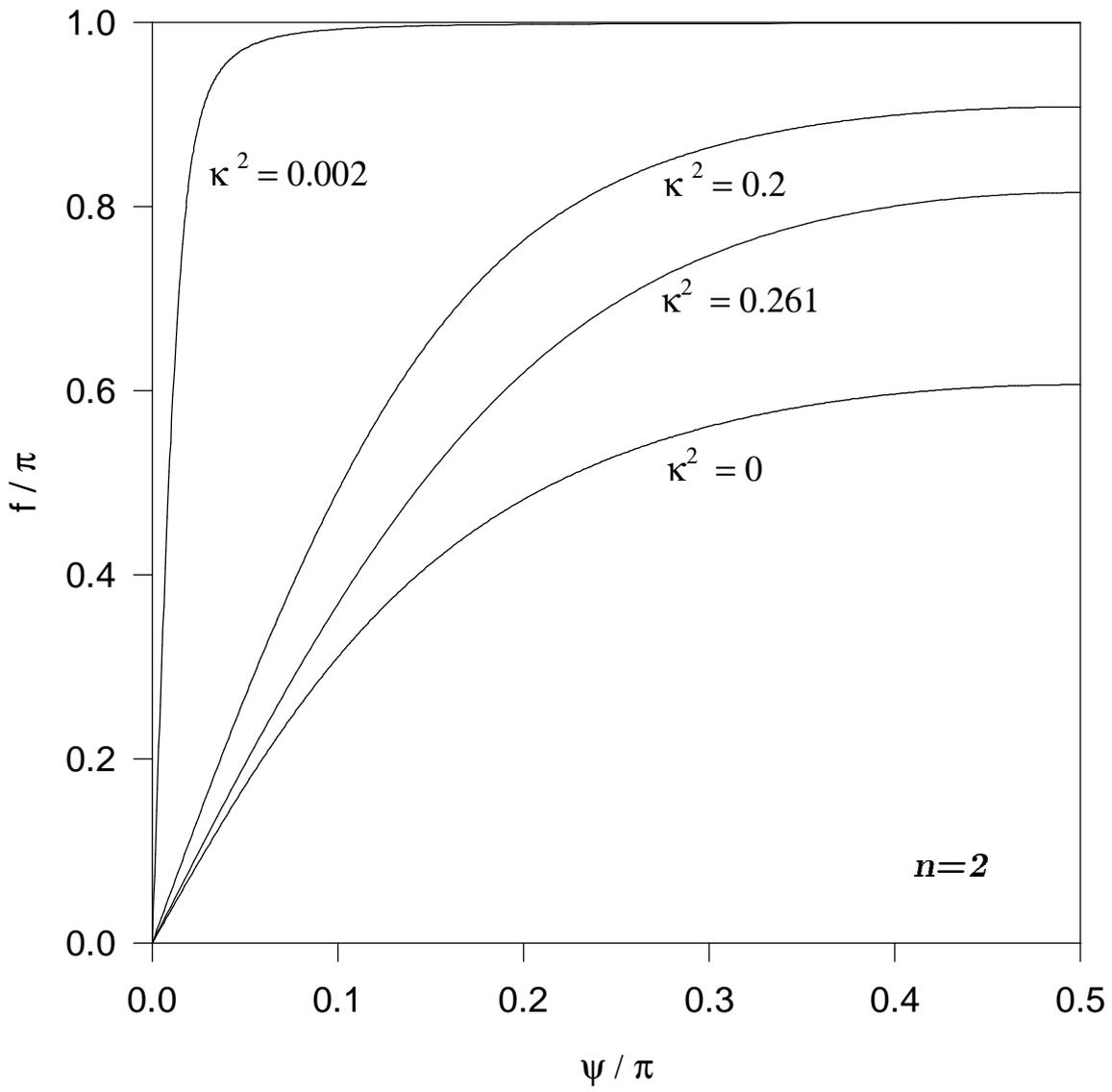
\item [Figure 1]
The solution $f_2(\psi,\kappa^2) / \pi$ is plotted in function of 
$\psi / \pi$  for four different values of the
coupling constant $\kappa$.
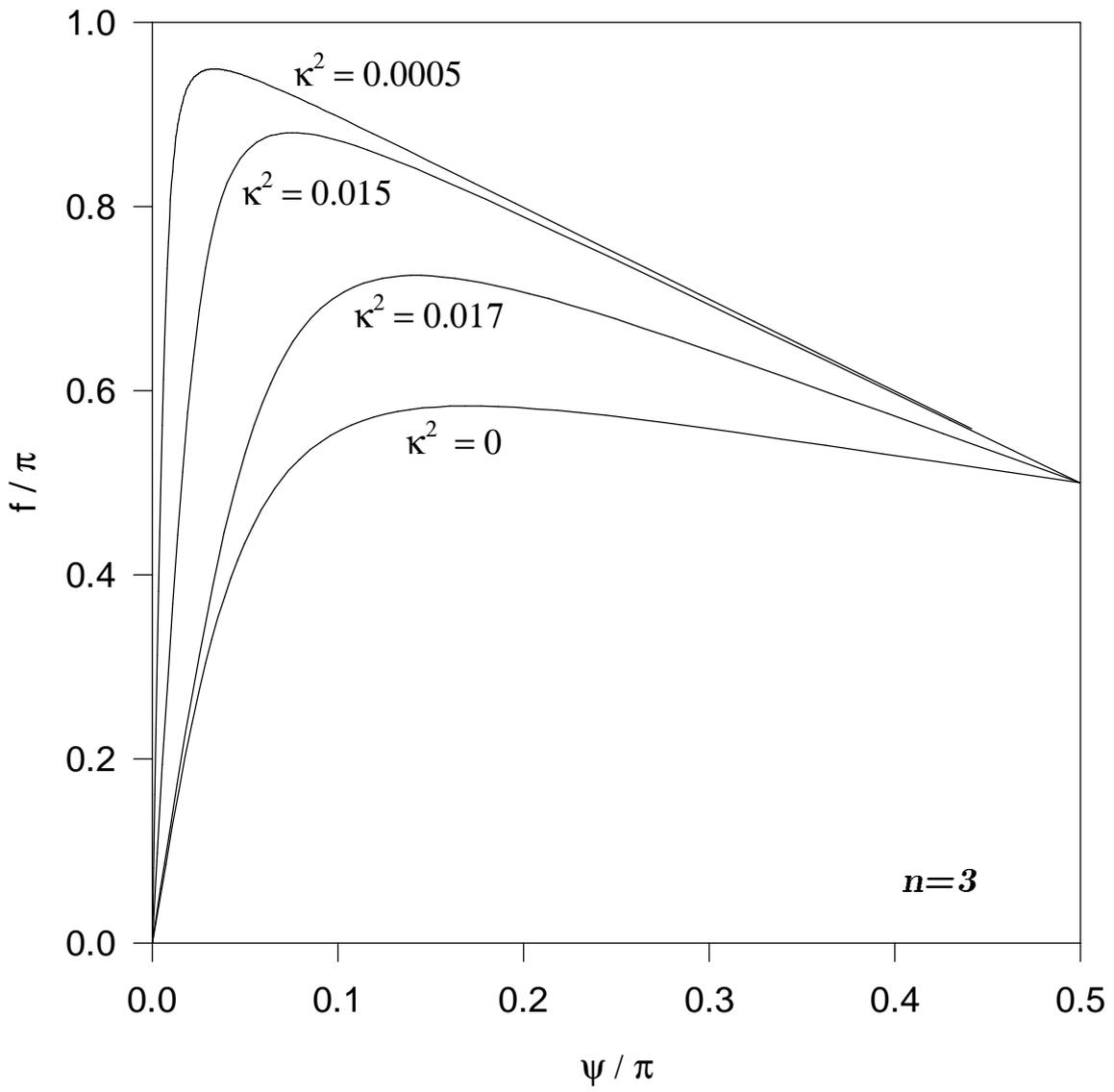
\item [Figure 2]
The solution $f_3(\psi,\kappa^2) / \pi$ is plotted in function of
$\psi / \pi$  for four different values of the
coupling constant $\kappa$.
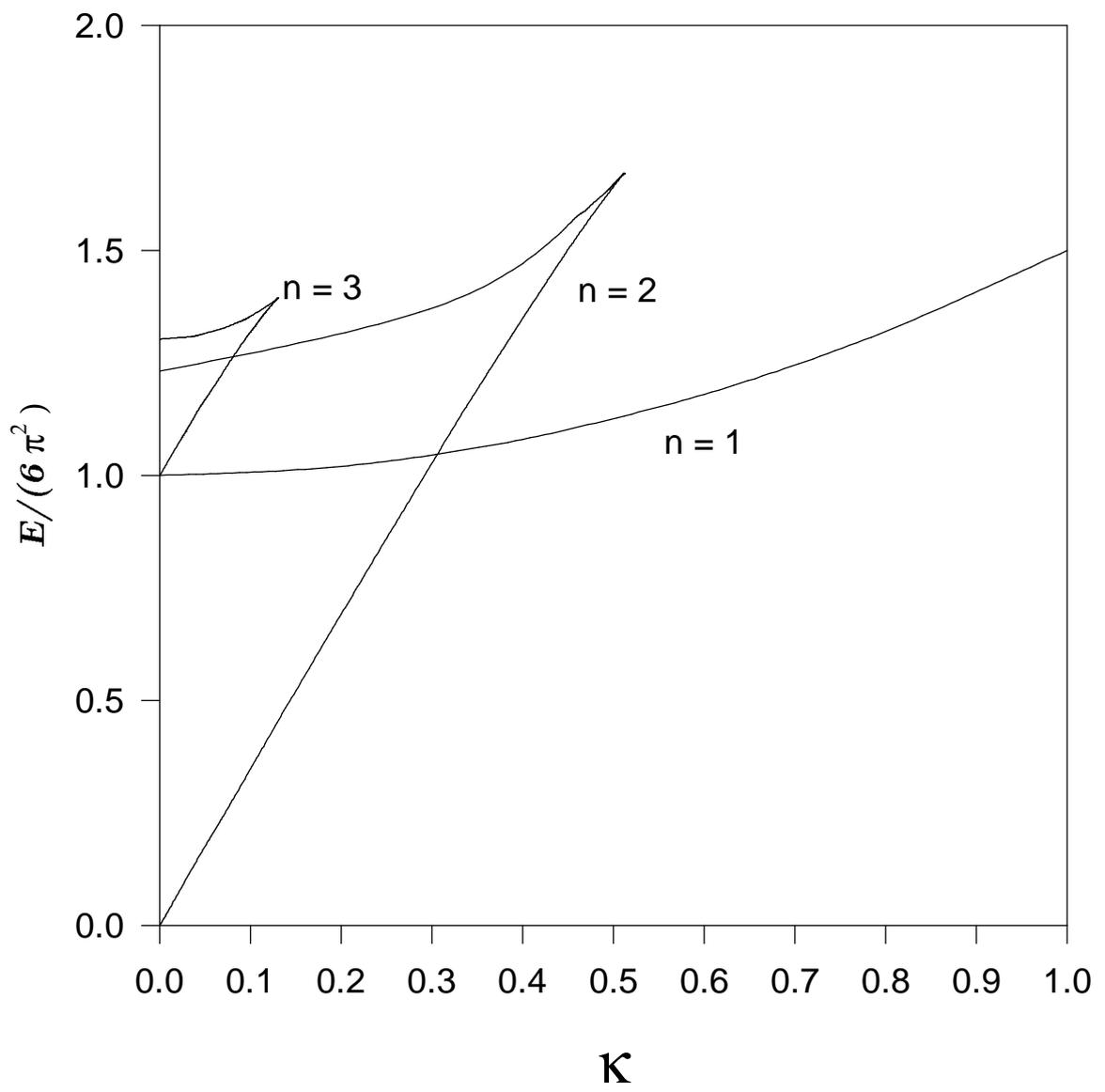
\item [Figure 3]
The energy of the solutions corresponding to $n=1,2,3$ is plotted
in function of the coupling constant $\kappa$. 
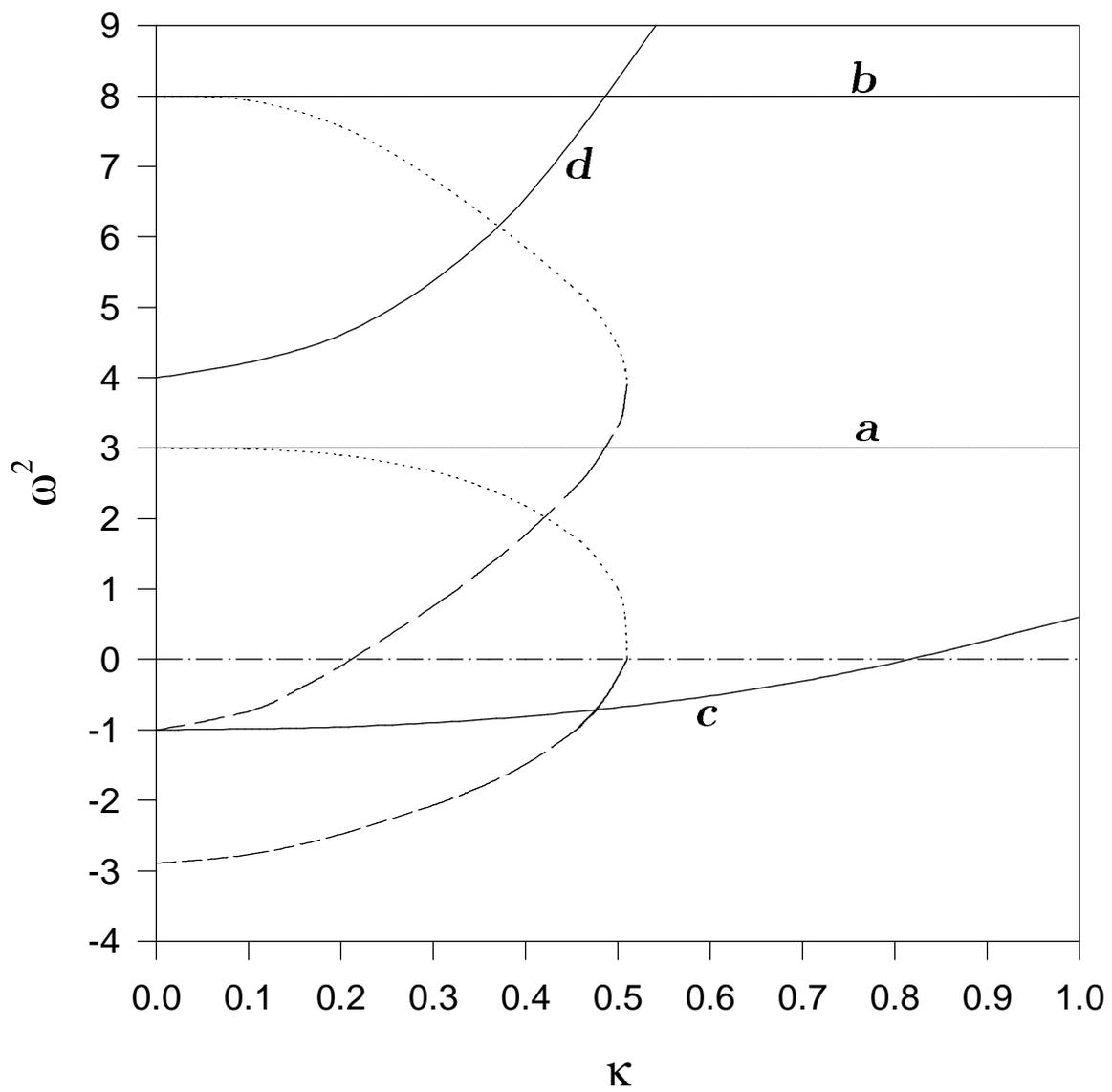
\item [Figure 4]
Some eigenvalues  of Eq.(\ref{mode}) are plotted in function
of $\kappa$. The lines $a,b$ refer to the vacuum (\ref{geg0})
and the lines $c,d$ to the $n=1$ solution.
The dashed (resp. dotted) lines refer to the modes of the
solution $f_2$ (resp. $\tilde f_2$).
\end{itemize}


\newpage
\begin{thebibliography}{99}
\bibitem{bi}
P. Bizon, {\it Harmonic maps between three-spheres},
hep-th/9407140.
\bibitem{sk}
T.H.R. Skyrme, Proc. Roy. Soc.{\bf A260} (1961) 127; Nucl.Phys. {31} (1962)
556.
\bibitem{boku}
J. Boguta and J. Kunz, Phys. Lett. {\bf B166} (1986) 93.
J. Kunz and D. Masak, Phys. Lett. {\bf B179} (1986) 176.
\bibitem{brku}
Y. Brihaye and J. Kunz, Mod. Phys. Lett. {\bf A4} (1989) 2723.
\bibitem{eks}
G. Eilam, D. Klabucar and A. Stern, Phys. Rev. Lett. {\bf 56} (1986) 1331.
\bibitem{bks}
Y. Brihaye, J. Kunz, C. Semay, Phys. Rev. {\bf B42} (1990) 2846.
\bibitem{jap} M. Bando et al. Phys. Rev. Lett. {\bf 54} (1985) 1215.
\end{thebibliography}
\end{document}